\documentclass[prl,superscriptaddress,twocolumn]{revtex4}
\usepackage{epsfig}
\usepackage{latexsym}
\usepackage{amsmath}
\begin {document}
\title {A model of student's dilemma}
\author{Adam Lipowski}
\affiliation{Faculty of Physics, Adam Mickiewicz University,
61-614 Pozna\'{n}, Poland}
\author{Ant\'onio L.~Ferreira}
\affiliation{Department of Physics, University of Aveiro, 3810
Aveiro, Portugal}
%%%%%%%%%%%%%%%%%%%%%%%%%%%%%%%%%%%%%%%%%%%%%%%%%%%%%%%%%%%%%%%%%%%%%%%%%%%%%%
\begin {abstract}
Each year perhaps millions of young people face the following
dilemma: should I continue my education or rather start working
with already acquired skills. Right decision must take into
account somebody's own abilities, accessibility to education
institutions, competition, and potential benefits. A multi-agent,
evolutionary model of this dilemma predicts a transition between
stratified and homogeneous phases, evolution that diminishes
fitness, fewer applicants per seat for decreased capacity of the
university, and presence of poor students at \'elite universities.
\end{abstract}
\maketitle
%%%%%%%%%%%%%%%%%%%%%%%%%%%%%%%%%%%%%%%%%%%%%%%%%%%%%%%%%%%%%%%%%%%%%%%%%%%%%
%%%%%%%%%%%%%%%%%%%%%%%%%%%%%%%%%%%%%%%%%%%%%%%%%%%%%%%%%%
Education system constitutes an important element of every modern
society. The main factor that attract people to education
institutions is the fact that a knowledgeable individual receives
a certain 'pay-off' that might translate into a better salary,
fame, or self-satisfaction. But to be educated and harvest
potential benefits, an individual must invest certain amount of
time and very often money. In addition, there is a risk related to
this process. For example, one can fail at the exam or loose
interest in an earlier chosen subject. Moreover, one has to consider that
only a fraction of a society can get access to education
institutions and competition is sometimes fierce. Thus, before
deciding whether to be, e.g., a university applicant, one has to take into
account the above factors as well as carefully evaluate his/her
own abilities.

Since abilities are difficult to assess and the number of
competitors is difficult to foresee, there is no a simple recipe
on making the right decision. As a result predicting the
functioning of an educational system is far from trivial. One
problem, for example, is a possibility that a strong pressure will
push all individuals toward being university applicants. Such a
situation might be quite frustrating for the society as a whole
since a large fraction of them is bound to fail. Perhaps it would
be better if part of a society would give up  educational
benefits and abstain from applying. But can we expect that optimal
decisions of individuals will imply a well-being of a society?
Another problem, is to predict the number of university
applicants. This number is of much concern not only to university
planners but also to applicants themselves. The later
ones are more concerned with the ratio of the number of applicants
and the amount of applicants the university will accept, since it
better reflects the actual competition. How are these numbers changing when
the university e.g., reduces the amount of accepted candidates?
Will this ratio, and at the same time competition, increase? Or, will
this imply a better level of university students? Naively, one can
expect that \'elite universities that accept only few best
applicants should be able to keep a very high level. However, a large
risk of failure might put off potential high-level applicants opening
the door to lower-level applicants. It seems that
modelling of educational systems so far is  restricted to certain
specific topics~\cite{Sklar}. Certainly, some insight into more basic
properties of such complex problems would be desirable that
hopefully could improve our educational systems.

In the present paper we introduce a multi-agent evolutionary model
that describes some aspects of educational systems. Such an
approach proved to be very successful in studying some other
conflicting social situations, known as the so-called prisoner
dilemma~\cite{pd}  providing thus an elegant game-theory
illustration of  reciprocal altruism~\cite{altruism}. In our model
there are $N$ individuals of levels of abilities $l_i$, and 'confidence
level' $p_i$, $i=1,\ldots, N$ ($0<l_i,p_i<1$). An elementary step (a unit of time) of the dynamics
of our model is made of four parts: decision making, payoff,
removal, and reproduction. First, each individual decides,
according to a stochastic rule specified below, whether to apply
or not to the university. The university accepts applicants
provided that their total number is not greater than $M$. If so,
only $M$ best (i.e., those with the highest values of $l_i$)
applicants are accepted. A successful applicant receives a payoff
$s$, while the rejected ones the failure payoff $f$. Those who
decided not to apply receive the give-up payoff $g$. In our
context it is natural to examine only the case $f<g<s$. The
received payoff determines fitness of an individual and thus its
survival and reproduction chances. More specifically, $K$
individuals are removed from the population according to the rule that a randomly chosen
individual $i$ that received a payoff $x_i(=f,g \ {\rm or}\  s)$
is removed with a probability $p(x_i)={\rm exp}(-x_i/z)$ where
$z=\frac{1}{N}\sum_{i=1}^{N}x_i$ is the average fitness of the
population. Finally, we create $K$ new individuals reproducing
surviving individuals. In particular, a selected survived
individual produces an offspring with probability $1-{\rm
exp}(-x_i/z)$. An offspring inherits parameters $l_i$ and $p_i$ of
his parent subject to a small mutation of amplitude $\delta$~\cite{mutation}.

A decision of $i$-th individual whether to apply to the university
or not is based on the comparison of his/her level $l_i$ with the
average level $\overline{l}$ of successful applicants of the
previous enrollment. The simplest rule might be for example to
apply to the university only when $l_i>\overline{l}$ and not
apply otherwise. However, we use another rule that allows for both
choices to be made albeit with some probability that is specified
by the 'confidence level' $p_i$. In particular, an individual
applies to the university with a probability $h(l_i,p_i)$ being a
continuous piece-wise linear increasing function of $l_i$ and $p_i$ defined as
\begin{equation}
h(l_i,p_i)=\left\{ \begin{array}{ll}
p_il_i/\overline{l} & {\rm for}\ l_i<\overline{l}\\
\big[(1-p_i)l_i-\overline{l}+p_i\big]/(1-\overline{l}) & {\rm for}
\ l_i\geq\overline{l}
\end{array}
\right. \label{eq1}
\end{equation}
The function $h(l_i,p_i)$ has the property that $h(l_i=0,p_i)=0$,
$h(l_i=1,p_i)=1$ and $h(l_i=\overline{l},p_i)=p_i$.

To examine the behaviour of our model we made numerical
simulations. Presented results are obtained for $N=1000$ but we
also made simulations for $N=500, 2000$ and $5000$ with
qualitatively the same behaviour. We used payoff values $s=10$,
$f=0$, the number of removed individuals $K=N/10$, and the
mutation amplitude $\delta=0.01$ (qualitatively the same behaviour
was seen for $\delta=0.001$). Varying the give-up payoff $g$ and the
normalized capacity of the university $m=M/N$, we monitored
certain characteristics of our model such as the ratio $r$ equal to the number of
individuals that decide to apply to the university divided by $N$, the
average fitness $z$, and some probability distributions. Unless
specified otherwise, simulations start with all individuals having
$l_i$ and $p_i$ randomly chosen from the unit interval $(0,1)$.
Stationary averages are calculated only after relaxing the system for sufficiently long time.
Obtained results are described below~\cite{applet}.
%%%%%%%%%%%%%%%%%%%%%%%%%%%%%%%%%%%%%%%%%%%%%%%%%%%%%%%%%%%%
\begin{figure}
\centerline{ \epsfxsize=9cm \epsfbox{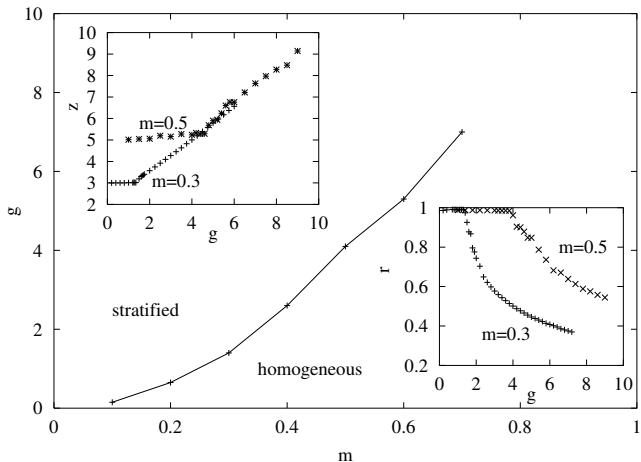} }
%\figspace
\caption{Transition line in the $(g,m)$ plane separating
homogeneous and stratified phases. Upper and lower insets show the
fitness $z$, and average number of individuals that decided to
apply to the university $r$, as functions of $g$.} \label{fig1}
\end{figure}
%%%%%%%%%%%%%%%%%%%%%%%%%%%%%%%%%%%%%%%%%%%%%%%%%%%%%%%%%%%%

First, numerical simulations show that the population with the
initial uniform distribution of $l_i$ and $p_i$ quickly evolves
toward much different states. Stationary behaviour, (see
Fig.~\ref{fig1}) depends on parameters $g$ and $m$. For
sufficiently small $g$ or sufficiently large $m$ the system
remains in the homogeneous phase where all individuals are
clustered around (1,1) point in the $(l_i,p_i)$ plane. In other
words, the population consists of individuals of a very good level
and which most likely will apply to the university ($r\simeq 1$).
Consequently, in this phase the average fitness remains very close
to the value $z=sm+f(1-m)=10m$ (i.e., a value when everyone is
applying), as  verified by numerical calculations (upper inset of
Fig.~\ref{fig1}).

When the give-up payoff $g$ is sufficiently large or the capacity
of the university $m$ is sufficiently small, struggling for
admission to the university  is no longer the best strategy.
Indeed, we observe that in such a case the model evolves
differently. Now, only a part of the population aims at the
university education and has $l_i$ and $p_i$ close to unity. In
addition to that there is a part of the population that receives
smaller but more secure give-up payoff $g$. To receive this payoff an individual should not apply
to the university and that is why in this part of the
population $l_i$ and $p_i$ are quite close to zero (let us notice that
according to Eq.~(\ref{eq1}) the smaller $l_i$ and $p_i$ are, the smaller
probability of applying to the university is). In this stratified
phase the ratio of applicants $r$ is definitely smaller than
unity and the fitness $z$ is larger than in the homogeneous phase
(insets of Fig.~\ref{fig1}).

As we already mentioned, the larger the fitness of an individual
the larger its survival and reproduction chances. One can thus
expect that a system will evolve toward states of maximal fitness.
Such an expectation originates partially from our experience with
thermodynamical systems where dynamics typically drives a system
toward a state that minimizes its free energy. But numerical
simulations show that evolutionary systems like our model do not
meet these expectations.
%%%%%%%%%%%%%%%%%%%%%%%%%%%%%%%%%%%%%%%%%%%%%%%%%%%%%%%%%%
\begin{figure}
\centerline{ \epsfxsize=8cm \epsfbox{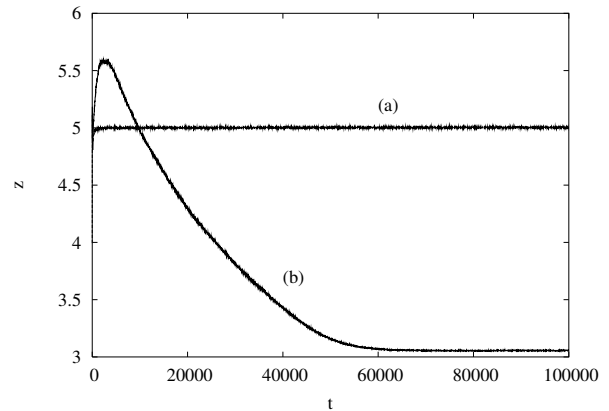} }
%\figspace
\caption{Time dependence of the global fitness $z$ for simulations
with the initial state such that (a) $0<p_i<1$ and (b)
$0<p_i<0.1$.
Simulations were done for $m=0.3$ and $g=4$ and results are averages of 100 independent runs.}
\label{time_fit}
\end{figure}
%%%%%%%%%%%%%%%%%%%%%%%%%%%%%%%%%%%%%%%%%%%%%%%%%%%%%%%%%%
In Fig.~\ref{time_fit} we show the time dependence of the fitness
$z$ calculated for $g=4$, $m=0.3$ and $N=1000$. For such values of
$g$ and $m$, when the system starts with $l_i$ and $p_i$ uniformly
drawn from the interval (0,1) it evolves toward the stratified
phase (see Fig.~\ref{fig1}). We checked that there is a lot of
other initial conditions (i.e., ways to draw initial values of
$l_i$ and $p_i$) that lead to the same phase. However, some
initial conditions yields a different behaviour. In particular,
when $p_i$ are initially drawn from a smaller interval namely,
(0,0.1), the system evolves  toward the homogeneous phase with all
$l_i$ and $p_i$ close to unity (Fig.~\ref{time_fit},
Fig.~\ref{config}). Qualitative understanding of such a behaviour
is as follows: When individuals have small values of $p_i$, the
number of individuals that apply to the university is also small,
and in particular smaller than the capacity of the university $M$.
As a result, nobody gets the failure payoff $f$ and the population
consists only of individuals with payoffs $s$ and $g$. Since $s>g$
individuals who apply to the university gain some evolutional
advantage. Of course, under such conditions individuals with
larger $p_i$  and $l_i$ are preferred since for them probability
to apply to the university is larger, and individuals with small
$p_i$ and $l_i$ are depleted. During such an evolution the
fraction of people applying to the university $r$ increases (that
leads to the increase of the average fitness $z$
(Fig.~\ref{time_fit})) and at a certain moment the population
reaches the point where $r=m$. At this point the population has
the largest possible fitness $z_M=sm+g(1-m)$ that in our case
$s=10,\ g=4$ and $m=0.3$ yields $z_M=5.8$~\cite{comment1}.
However, the evolution does not stop here but drives the system
towards the homogeneous phase with much smaller fitness $z$
(Fig.~\ref{time_fit}). Let us also notice that even in the
stratified phase the fitness $z(\sim 5.0)$ is lower (although not
that much) than the maximal value $z_M$.
%%%%%%%%%%%%%%%%%%%%%%%%%%%%%%%%%%%%%%%%%%%%%%%%%%%%%%%%%%
\begin{figure}
\centerline{ \epsfxsize=9cm \epsfbox{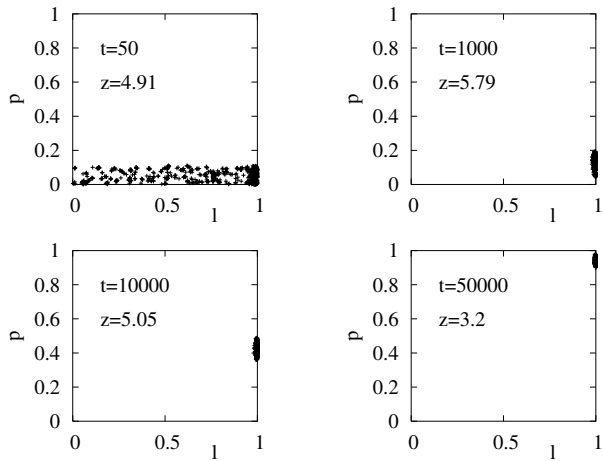} }
%\figspace
\caption{Snapshots of distributions of $l_i$ and $p_i$ for
simulations with the initial state such that $0<p_i<0.1$.
Simulations were done for $m$ and $g$ as in Fig.~\ref{time_fit}.
For $t=1000$ the system is very close to the state with maximal
fitness $z=5.8$} \label{config}
\end{figure}
%%%%%%%%%%%%%%%%%%%%%%%%%%%%%%%%%%%%%%%%%%%%%%%%%%%%%%%%%%%%

\begin{figure}
\centerline{ \epsfxsize=8cm \epsfbox{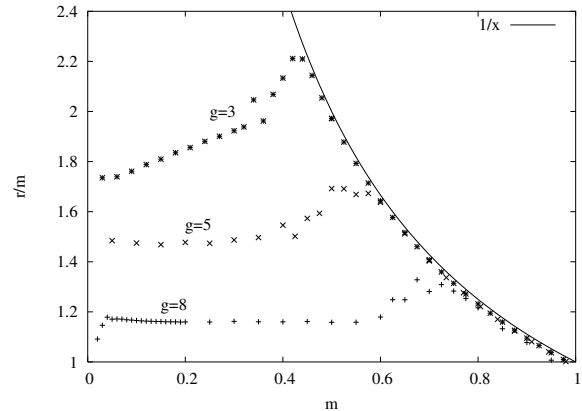} }
%\figspace
\caption{The number of applicants per seat as a function of the
capacity of the university $m$ calculated for several values of
$g$.} \label{ilezd}
\end{figure}

An important characteristic that can be extracted from our model
is an average number of applicants $r$ divided by the capacity of
the university $m$. This is actually the number of applicants per
seat and it is an obvious measure of competition for the
admission. In Fig.~\ref{ilezd} we present numerical calculation of
this quantity.
%%%%%%%%%%%%%%%%%%%%%%%%%%%%%%%%%%%%%%%%%%%%%%%%%%%%%%%%%%
%%%%%%%%%%%%%%%%%%%%%%%%%%%%%%%%%%%%%%%%%%%%%%%%%%%%%%%%%%%%

First, one can see that in the homogeneous phase $r/m$ is well
approximated with the function $1/m$, which is an straightforward
consequence of the fact that in this phase probability of applying
to the university is very large and hence $r$ is very close to
unity. The behaviour dramatically changes in the stratified phase
where the number of applicants per seat $r/m$ decreases when the
capacity $m$ decreases. Apparently, a decreased capacity makes an
admission less likely and that discourages potential applicants.
Moreover, one can see that a maximum of $r/m$ that corresponds to
the most fierce competition among candidates occurs most likely at
the transition between stratified and homogeneous phases.

\begin{figure}
\centerline{ \epsfxsize=9cm \epsfbox{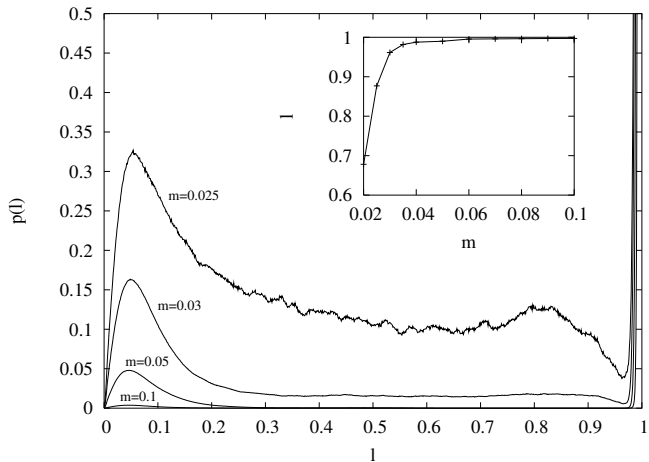} }
%\figspace
\caption{Probability distributions for accepted individuals to
have the level $l$ calculated for $g=8$ and several values of $m$.
Inset shows the average level $\overline{l}$ as a function of $m$
for $g=8$} \label{probd}
\end{figure}

Although the number of applicants per seat decreases, for $g=3$
and 5 it is still definitely larger than unity. As a result the
university has a plenty of candidates to chose from, and the level
of university students $\overline{l}$ is very close to unity. An
indication of a possibly different behaviour is seen in
Fig.~\ref{ilezd} for $g=8$ and very small $m$ (where a sharp drop of$r/m$ is seen).
Indeed, in this case, the average level $\overline{l}$
is definitely smaller than unity (inset of Fig.~\ref{probd}). To
explain such a behaviour we looked in more details at individuals
accepted to the university. It turns out that for small $m$ a non
negligible fraction of successful applicants has quite a low level
$l_i$ (Fig.~\ref{probd}). Let us recall that for such values of
$g$ and $m$ the system is in the stratified phase. Typically, even
in the stratified phase successful applicants belong only to that
part of the population where $l_i$ and $p_i$ are very close to
unity. However, Fig.~\ref{probd} shows that for $g$ close to $s$
and small $m$ the other part of the population (i.e., that with
$l_i$ and $p_i$ close to zero) gets access to the university as
well. Although it reduces the level of applicants $\overline{l}$,
in a wider social context it might be a desirable feature that
candidates from other (less educated) group can enter such an
\'elite university.
%%%%%%%%%%%%%%%%%%%%%%%%%%%%%%%%%%%%%%%%%%%%%%%%%%%%%%%%%%
%%%%%%%%%%%%%%%%%%%%%%%%%%%%%%%%%%%%%%%%%%%%%%%%%%%%%%%%%%%%%%%

Finally, let us notice that certain features of our model bear
some similarity to minority games~\cite{minority}. Indeed, the
payoff of an individual in our model depends on the behaviour of
other individuals. Moreover, individuals have a one-step memory
and remember the average level $\overline{l}$ of successful
applicants from the previous enrollment. Although our model seems
to be more complicated, we hope that it might be interesting to
examine further similarities with minority games since the later
ones were already subject of intensive research and are  much
better understood. For example on can study a version of our model
where individuals have a longer memory and see how performance of
an individual depends on its memory. Another possible
generalization might be to examine the case of two universities
(with e.g., different success payoffs $s$) and/or the role of
spatial effects. Such effects are known to play an important role
in the evolutionary versions of prisoner's dilemma
model~\cite{nowak}.

Nevertheless, the behaviour of our model seems to be to some extent generic. Indeed, simulating the model with probability
of applying to the university being independent of $l_i$ and given as
$h(l_i,p_i)=p_i$ we also observed e.g., the existence of
stratified and homogeneous phases. However, such a choice of
$h(l_i,p_i)$ we consider as less realistic. Moreover, these phases appear also for all $p_i$ equal to unity (and kept constant). However, some other effects reported in this paper do not appear in this case.

In summary we introduced an evolutionary, multi-agent model of an
enrollment process, where a population of individuals of different
levels of abilities $l_i$ and confidence $p_i$ have to decide
whether to apply or not to the university. Depending on their
decisions (and decisions of all other individuals) they receive
certain payoffs. Such a payoff determines fitness of a given
individual and that in turn determines its chance of survival and
reproductivity. Numerical simulations show that depending on
parameters (payoff values or capacity of the university) the
system might be in either a homogeneous phase, with all
individuals having $l_i$ and $p_i$ close to unity (i.e., taking
almost maximal values ) or in the stratified phase where a part of
the population has much lower values of $l_i$ and $p_i$. Although
dynamics of the model favors individuals with a large fitness
under certain conditions the model is driven toward states where
the global fitness of all individuals is quite low. Even more, on
the way to such states it passes through states of large fitness,
that unfortunately cannot trap it. This result is actually quite
worrying and perhaps not unrealistic. It shows that a society
might be driven toward an undesirable state even though
individuals are making no explicit efforts to enter such a state.
Moreover, our simulations show that decreasing the capacity of the
university the number of candidates per seat increases but only in
the homogeneous phase. When this capacity is too low the model
enters a stratified phase and the number of candidates per seat
decreases (apparently put off by a large risk of failure). In the
extreme case of a very small capacity, competition drops even
further and as a result applicants of a quite poor abilities (from
a different social group) can succeed.

Acknowledgements: The research grant 1 P03B 014 27 from KBN is
gratefully acknowledged. Numerical calculations were partially
performed on ' Open Mosix Cluster' built and administrated by
dr.~L.~D\c{e}bski.
%%%%%%%%%%%%%%%%%%%%%%%%%%%%%%%%%%%%%%%%%%%%%%%%%%%%%%%%%%%%

%%%%%%%%%%%%%%%%%%%%%%%%%%%%%%%%%%%%%%%%%%%%%%%%%%%%%%%%%%%%%%%%%%%%%%%%%%%%%%%
%%%%%%%%%%%%%%%%%%%%%%%%%%%%%%%%%%%%%%%%%%%%%%%%%%%%%%%%%%%%%%%%%%%%%%%%%%%%%%%

\begin{thebibliography}{}
\bibitem{mutation} Parameters $l_i$ and $p_i$ of an offspring are given as $l_i=l_i^p+\omega$, $p_i=p_i^p+\omega$, where
$\omega$ is a randomly chosen number such that
$-\delta/2<\omega<\delta/2$ and $l_i^p$ and $p_i^p$ are parent
parameters.
\bibitem{comment1} Since the point where $r=m$ for different runs is reached at
slightly different times, after averaging we obtain in
Fig.~\ref{time_fit} a peak of $z$ at a slightly smaller value than 5.8.
\end{thebibliography}
\end {document}